\documentclass[10pt]{ismd08}
\usepackage{graphicx}
\usepackage{sidecap} 

\usepackage{cite,./mcite}

\def\corr{\langle \Delta p_{t,1}, \Delta p_{t,2}\rangle}

\begin{document}
 \title{$D\overline{D}$ momentum correlations versus relative
azimuth as a sensitive probe for thermalization}
\author{G.~Tsiledakis\protect\footnote{\ \ speaker} and K.~Schweda}
\institute{University of Heidelberg Physikalisches Institut, D-69120 Heidelberg, Germany }
\maketitle
\begin{abstract}
In high-energy nuclear collisions at LHC, where a QGP might be created,
the degree of thermalization at the partonic level is a key issue. 
Due to their large mass, heavy quarks are a powerful tool to probe
thermalization. We propose to measure azimuthal correlations of heavy-quark hadrons and their decay
products.  Changes or even the complete absence of  these initially existing azimuthal correlations in $Pb-Pb$
collisions might indicate thermalization at the partonic level. We present studies with PYTHIA for
$p-p$ collisions at 14 TeV using the two-particle transverse momentum correlator 
${\langle\overline{\Delta}p_{t,1}\overline{\Delta}p_{t,2}\rangle}$
 as a sensitive measure of potential changes in these azimuthal
correlations. Contributions  from transverse radial flow are estimated.
%
\end{abstract}
%
\section{Introduction}
Ultra-relativistic heavy ion collisions offer the unique opportunity to probe highly
excited (dense) nuclear matter under controlled laboratory conditions. The compelling
driving force for such studies is the expectation that at high
enough temperature and/or density hadrons dissolve into a new form of elementary
particle matter, the Quark Gluon Plasma (QGP), where quarks and gluons
are deconfined.
An essential difference between elementary particle collisions
and nuclear collisions is the development of collective motion in the
latter. The collective flow of all hadrons, especially the multistrange hadrons $\phi$
and $\Omega$, has been experimentally measured  \cite{strange} at RHIC and suggest that collective motion
develops in the early partonic stage. Presently, the degree of
thermalization at the parton level is a crucial issue.

The observables related to heavy-quark hadrons are of particular interest in the study of
thermalization \cite{thermalization}. Heavy quarks remain massive in a QGP and can only be pair-created
in the early stage of the collisions contrary to light quarks which obtain their small bare masses in the
deconfined phase when chiral symmetry is partially restored. In the subsequent
evolution of the medium, the number of heavy quarks is conserved because the typical
temperature of the medium is much smaller than the thresholds for thermal heavy quark $(c, b)$
production. These heavy quarks participate in collective motion provided their
interactions at the partonic level occur at high frequency. Thus, collective motion
of heavy-quark hadrons will be a useful tool for studying the early thermalization of
light quarks in high-energy nuclear collisions.
\begin{SCfigure}
\centering
\vspace{0.1cm} 
\includegraphics[width=8cm]{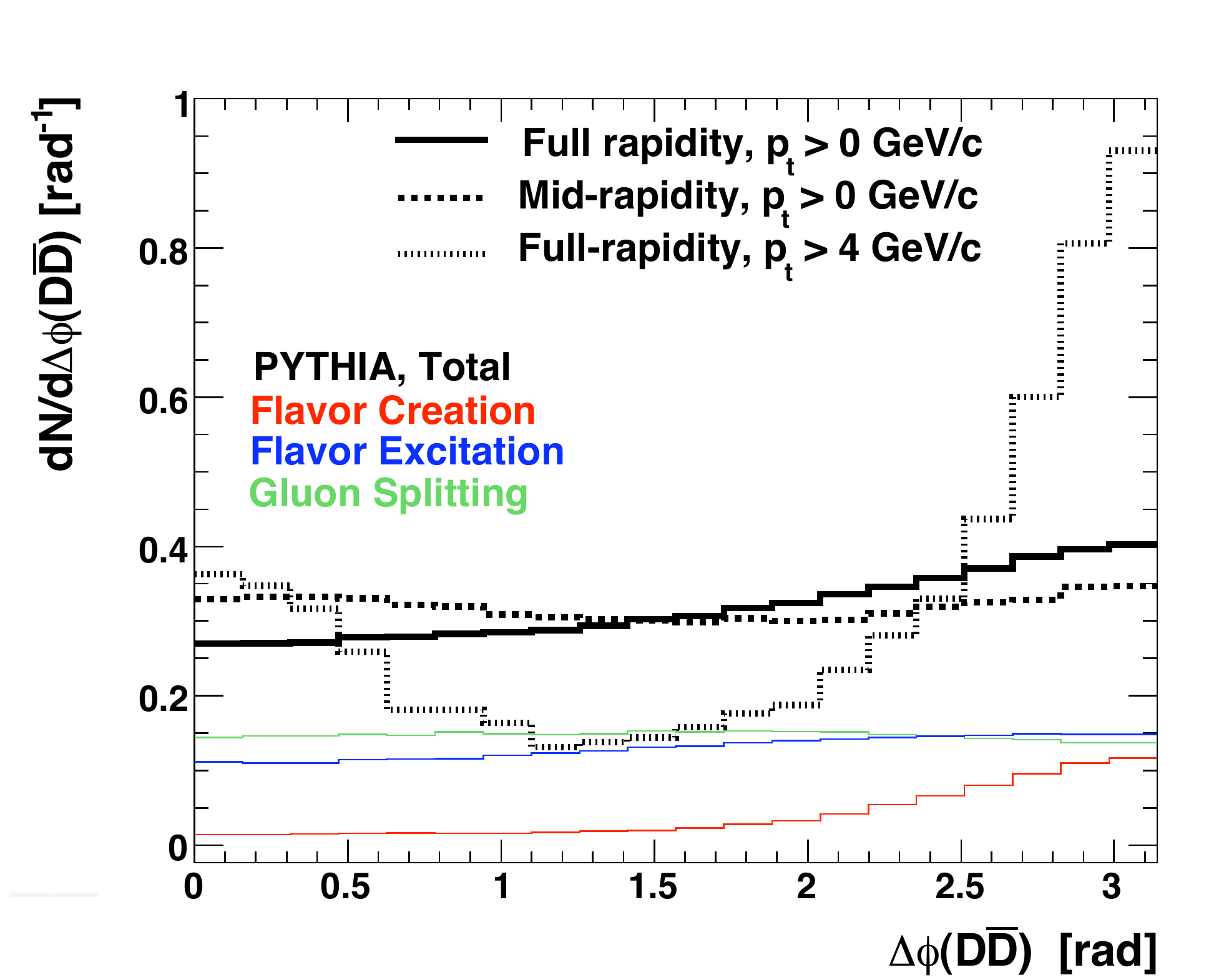}
\caption{(Colour online) Distribution in relative azimuth $\Delta\phi$ of $D\overline{D}$ pairs 
from $p-p$ collisions at $\sqrt{s}$ = 14 TeV as calculated by PYTHIA (v. 6.406), at full rapidity (solid line),
and mid-rapidity (dashed line). Contributions from flavor creation, flavor excitation and gluon
splitting to the full rapidity distribution are also shown as well as the
$p_{t}$ $>$ 4 GeV/c range (dotted line). }
\label{dndphi}
\end{SCfigure}

\subsection{$D\overline{D}$ angular correlations}
Since heavy quarks are pair-created by initial hard scattering processes, each
quark-antiquark pair is correlated in relative azimuth $\Delta\phi$ due to momentum
conservation. In elementary collisions, these correlations survive the fragmentation
process to a large extent and hence are observable in the distribution of $D$ and $\overline{D}$ mesons. 
 The observation of broadened angular correlations of
heavy-quark hadron pairs in high-energy heavy-ion collisions would be an
indication of thermalization at the partonic stage (among light quarks and
gluons) since the hadronic interactions at a late stage in the collision
evolution cannot significantly disturb the azimuthal correlation of 
$D\overline{D}$ pairs \cite{Zhu}. As a result, a visible decrease or the complete absence
of such correlations, would indicate frequent interactions of heavy quarks 
and other light partons in the partonic stage 
in nucleus-nucleus collisions at RHIC and LHC energies.

Concerning $p-p$ collisions, the Monte
Carlo event generator PYTHIA \cite{pythia} reproduces well the experimentally observed correlations of
$D$ mesons, measured at fixed target energies \cite{review}. Fig.~\ref{dndphi} shows the
calculated correlation for $p-p$ collisions at LHC energies ($\sqrt{s}$ = 14 TeV) where 
the PYTHIA (v. 6.406) parameters were tuned to reproduce the NLO 
predictions \cite{pythia2,Dainese} (with the option MSEL=1). The calculations at leading order (LO) which contain only flavor
creation processes ($q\overline{q}\rightarrow Q\overline{Q}$, $gg\rightarrow Q\overline{Q}$)
lead to back-to-back $D\overline{D}$ pairs. 
	Next-to-leading order (NLO) contributions like flavor excitation
($qQ\rightarrow qQ, gQ\rightarrow gQ$) and gluon splitting ($g\rightarrow Q\overline{Q}$) 
which become dominant at high energy, do not show explicit angular 
correlation leading to a strongly suppressed back-to-back correlation.
 At mid-rapidity, the $D\overline{D}$ correlation in $p-p$ collisions
at LHC energies has a rather flat angular distribution  \cite{Zhu2}. 
Thus, the measurement of these correlations and their modifications in $Pb-Pb$ collisions is challenging.
We introduce the two-particle transverse momentum correlator as a sensitive measure of heavy-quark correlations.  
  
\section{Employing the two-particle transverse momentum correlator}
The strong transverse momentum dependence of the $D\overline{D}$ correlation 
leading to a $\Delta\phi$ distribution peaked at 180$^{\rm o}$ for high $p_{t}$ $D$ mesons,
as one would expect for back-to-back pairs stemming from hard scatterings of partons
(Fig.~\ref{dndphi}).
For this purpose, an additional measure is introduced.
The occurence of non-statistical fluctuations of the event-by-event mean 
transverse momentum $M_{pt}$ goes along 
with correlations among the transverse momenta of particles. 
Such correlations can be calculated
employing the two-particle transverse momentum correlator \cite{pipj,ceres-pt} 
 for $D$ and $\overline{D}$ respectively.
\begin{equation}
\corr^{(D\overline{D})}=
\frac{1} 
{\sum_{k=1}^{n_{\rm ev}}N_k^{\rm pairs}}.
C_{k}
\end{equation}
where $C_{k}$ is the $p_{t}$ convariance:
\begin{equation}
C_{k}=
\sum_{i=1}^{N_k}\sum_{j=1}^{N_k}(p_{ti}-\overline{p_t}^{(D)})
(p_{tj}-\overline{p_t}^{(\overline{D})}) 
\end{equation}
where $p_{ti}$ and $p_{tj}$ are the $p_{t}$ for $i^{th}$ and $j^{th}$ 
track in an event of $D$ and $\overline{D}$ respectively,
$\overline{p_t}$ is the inclusive mean transverse momentum 
of all tracks from all events of $D$ and $\overline{D}$, 
 $\sum_{k=1}^{n_{\rm ev}}N_k^{\rm pairs}$ the total number of 
$D\overline{D}$ pairs and $n_{ev}$ the total number of $p-p$ collisions.
%
\begin{figure}[!th]
    \begin{center}
	\includegraphics[width=0.49\textwidth]{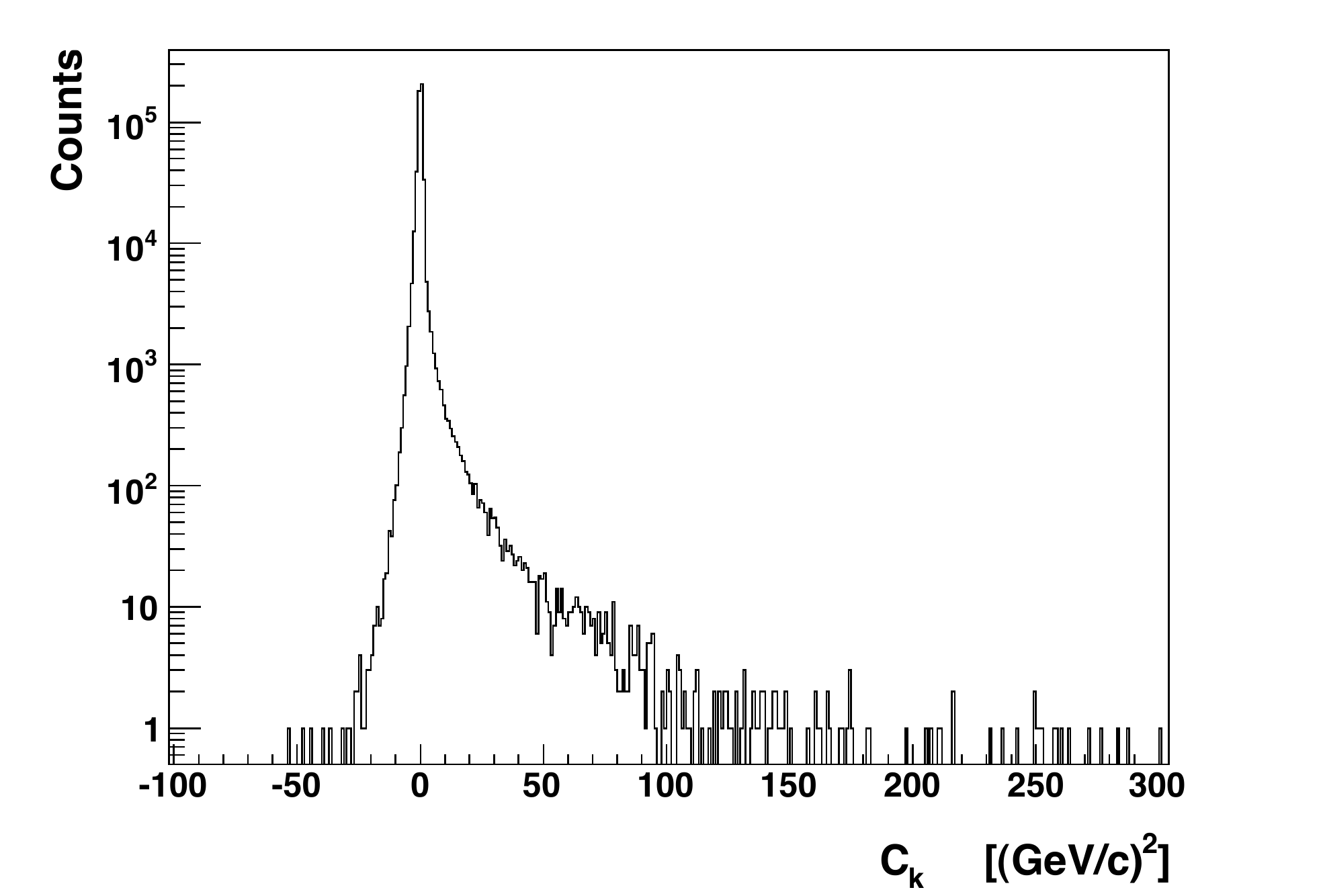}%
	\hspace{0.1cm}%
	\includegraphics[width=0.49\textwidth]{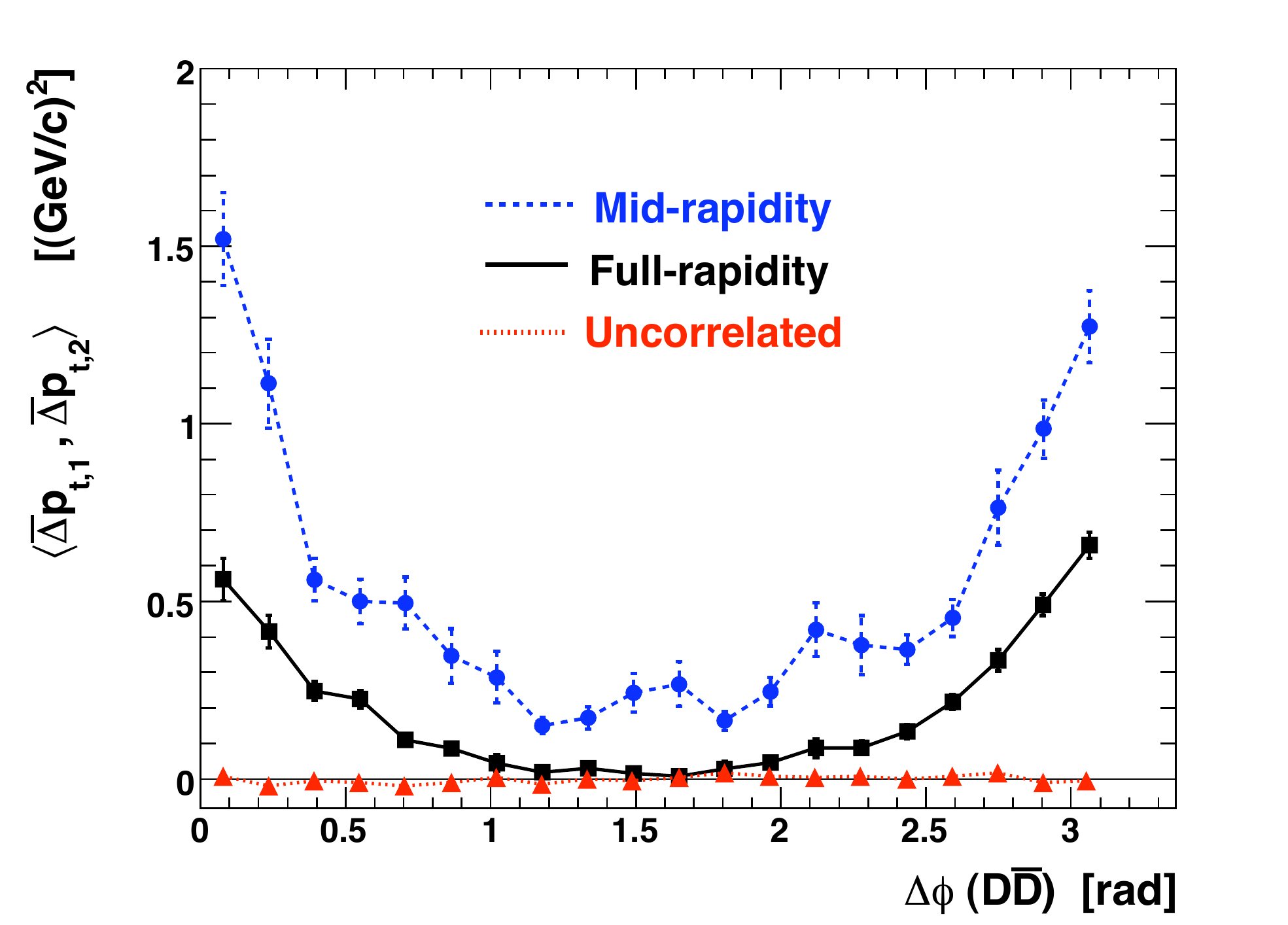}
    \end{center}
\caption{(Colour online) Distribution of the $p_{t}$ convariance $C_{k}$ for $D\overline{D}$ pairs 
from $p-p$ collisions at $\sqrt{s}$ = 14 TeV as calculated by PYTHIA (v. 6.406)
(left panel) and their momentum correlator $\corr$ as a function of $\Delta\phi$ 
at full rapidity, mid-rapidity and for background using the mixed event method (right panel).} 
\label{Ckpipj}
\end{figure}
It is also possible to study the scale dependence of $p_{t}$ correlations in
azimuthal space by calculating the correlator in bins of the azimuthal separation 
$\Delta\phi$ of particle pairs.
For the case of independent particle emission from a single parent distribution, $\corr$
vanishes. Fig.~\ref{Ckpipj} (left panel) shows the distribution of the $p_{t}$ convariance 
$C_{k}$ for $D\overline{D}$ pairs. The mean of this distribution reveals a strong correlation
with $\corr = 0.199\pm0.006$~GeV$^2$/$c^2$, which corresponds to the normalized dynamical 
fluctuation $\Sigma_{pt}$ \cite{spt} of $\sim30\%$ in $\overline{p_t}$. This is a strong
correlation when compared to 
$\sim1\%$ that was measured for unidentified charged particles in central collisions at SPS and RHIC \cite{spt,spt1,spt2}. 
The $D\overline{D}$ momentum correlator $\corr$ as a function of relative
azimuth angle $\Delta\phi$ is shown in Fig.~\ref{Ckpipj} (right panel).
Using particles from different $p-p$ collisions, which are physically uncorrelated (mixed event method), results in a
value of $\corr$ consistent with zero, as expected.
Applying the correlator to $D\overline{D}$ mesons from the same $p-p$ collision,
 a rich structure is observed.
At full rapidity, the
most pronounced features are a strong peak at small angles due to gluon
splitting, while flavor creation of $c\overline{c}$-quark pairs results in a peak of similar magnitude at 
large angles. 
At mid-rapidity the signal is even enhanced, due to the harder $p_{t}$ spectrum
of $D\overline{D}$ meson-pairs. 

Our tests show that a realistic amount of elliptic flow does not change these
correlations. Concerning the radial flow contribution, it is assumed that the 
expansion produces an additional momentum $p_{t,f}=\gamma m \beta$, where $\gamma$ is the Lorentz
factor, $\beta$ is the profile velocity and $m$ is the mass of the $D$ meson.
By adding this radial flow component \cite{flow} vectorially to the momentum vector produced by PYTHIA, we 
evaluate the $\corr$ as a function of $\Delta\phi$. 
\begin{SCfigure}
\centering
\vspace{0.cm} 
\includegraphics[width=7cm]{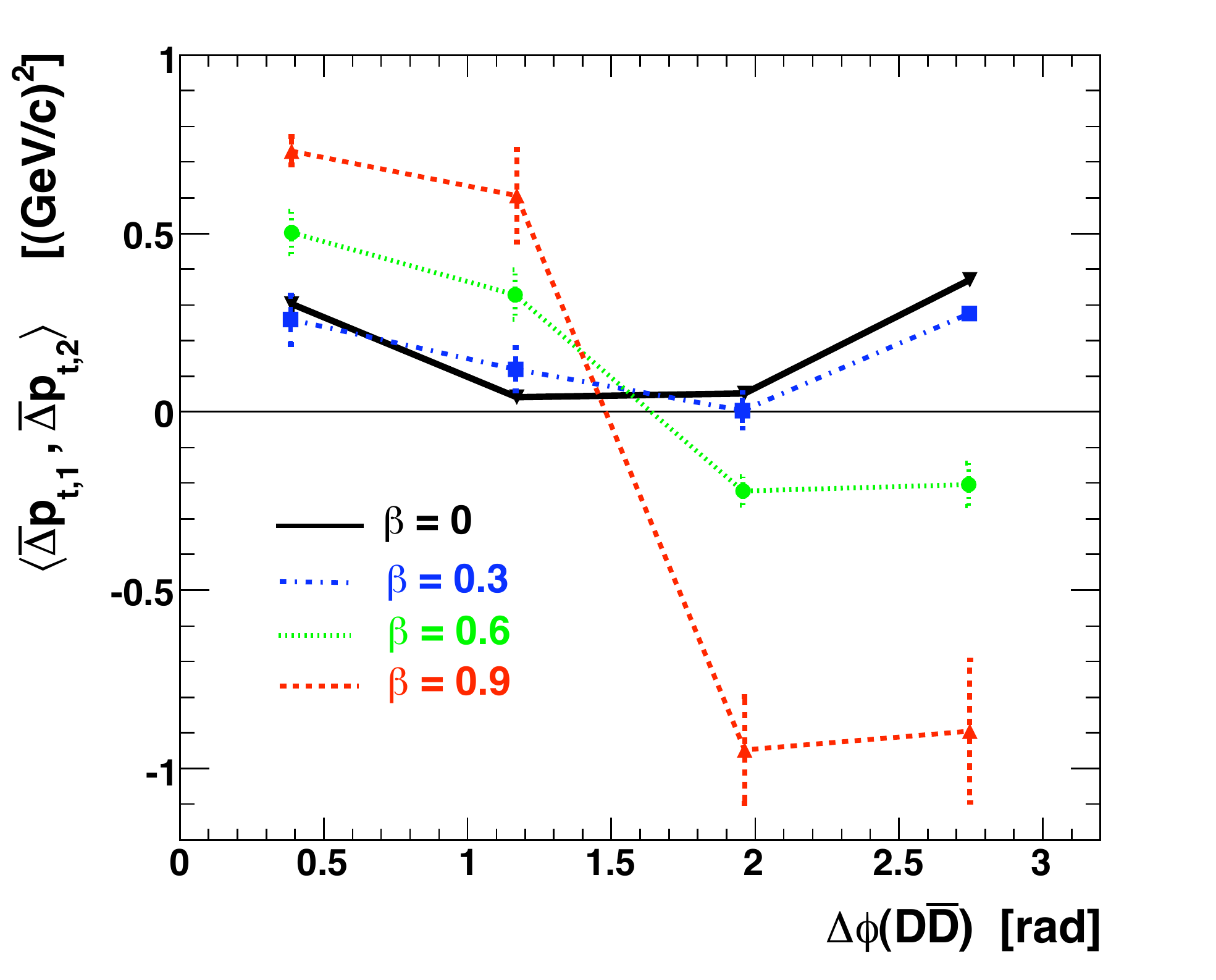}
 \caption{(Colour online) The momentum correlator $\corr$ of $D\overline{D}$ pairs 
 from $p-p$ collisions at $\sqrt{s}$ = 14 TeV as calculated by PYTHIA (v. 6.406),
 as a function
 of their relative azimuth $\Delta\phi$ 
 at full rapidity for various collective flow velocities $\beta$.}
\label{radial}
\end{SCfigure}
As it is shown in Fig.~\ref{radial}, strong radial flow will further increase
the same side momentum correlations of $D\overline{D}$ pairs and might lead to strong 
anti-correlations at large angles.

\begin{figure}[!th]
    \begin{center}
	\includegraphics[width=0.49\textwidth]{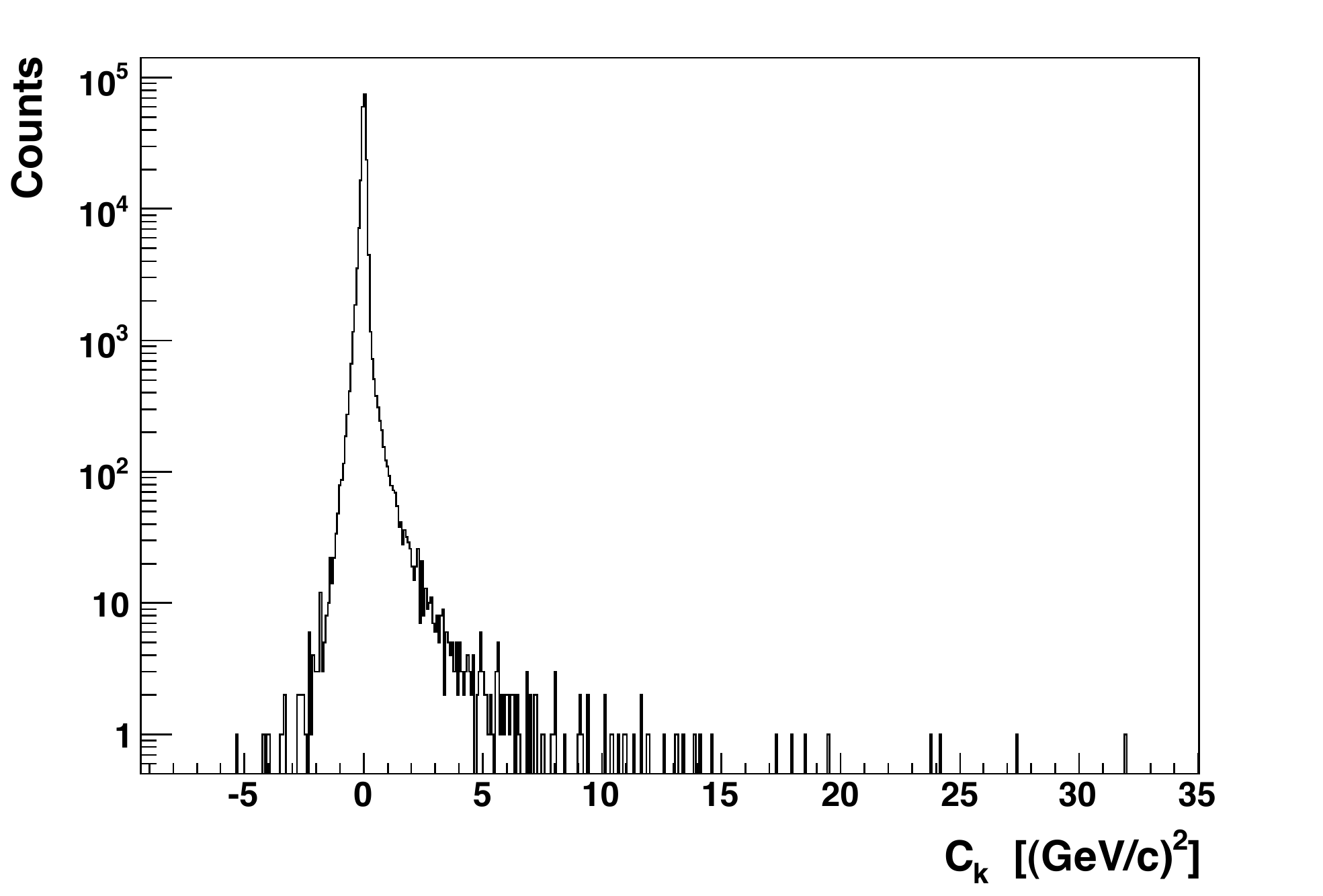}%
	\hspace{0.1cm}%
	\includegraphics[width=0.49\textwidth]{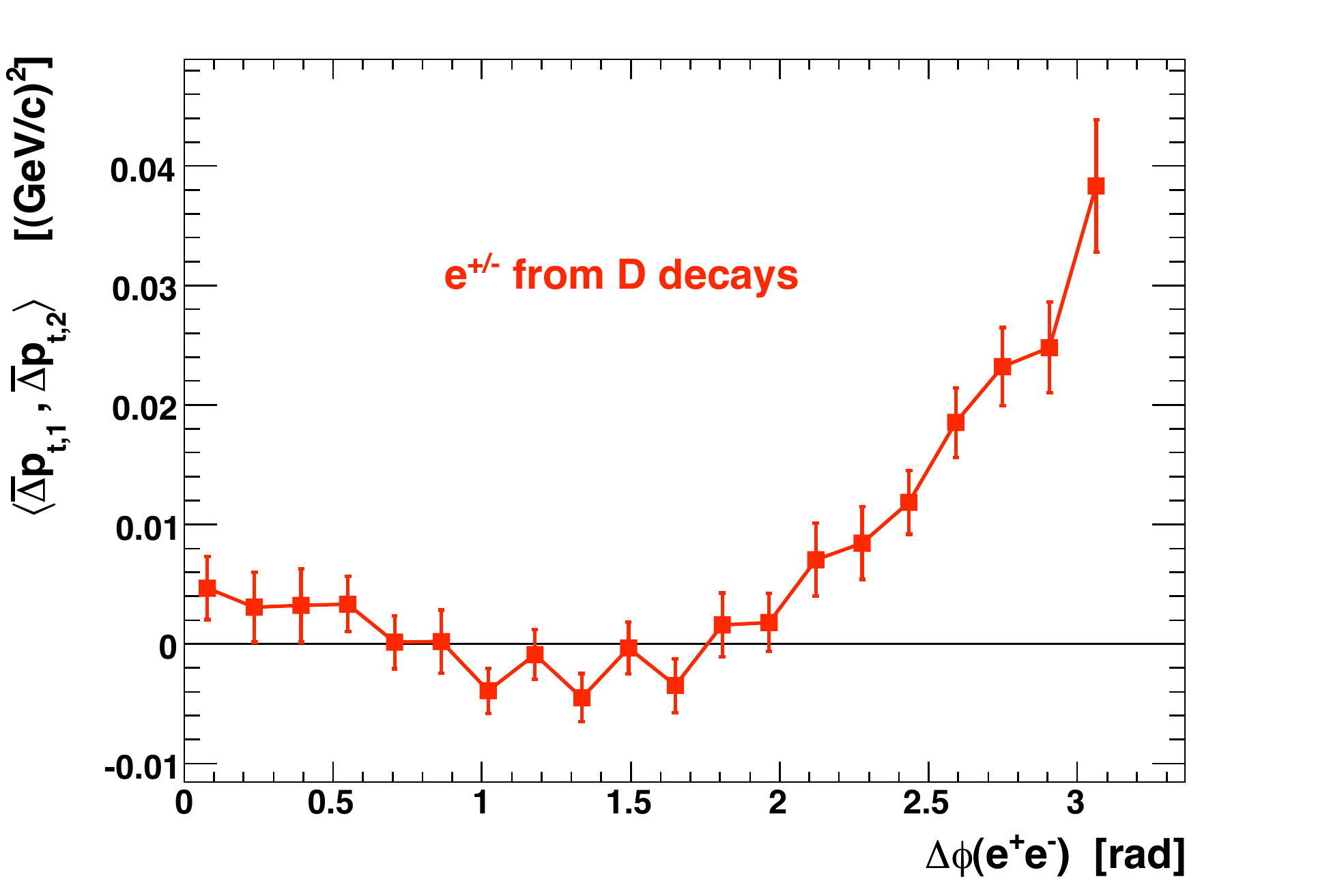}
    \end{center}
\caption{(Colour online) Distribution of the $p_{t}$ convariance $C_{k}$ for $e^{+}e^{-}$ pairs from $D\overline{D}$
decays from $p-p$ collisions at $\sqrt{s}$ = 14 TeV as calculated by PYTHIA (v. 6.406) 
(left panel) and their momentum correlator $\corr$ as a function of $\Delta\phi$ 
at full rapidity (right panel).} 
\label{elec}
\end{figure}

As we have shown, the initial correlations of $c\overline{c}$ pairs survive the fragmentation process. 
 However, direct reconstruction of $D$ mesons from topological decays suffer from small efficencies resulting in low statistics.
 Therefore, we investigated 
 semileptonic decays of $D$ mesons and performed an analogous analysis.
Our results indicate that dileptons from $D\overline{D}$ decay
preserve the original $D\overline{D}$ angular $p_{t}$ correlation to a large
extent. Fig.~\ref{elec} shows the momentum correlator $\corr$ of $e^{+}e^{-}$ pairs from $D\overline{D}$
decay, as a function of of their relative azimuth $\Delta\phi$ 
at full rapidity, where the away-side peak at large angles is sizeable
(right panel). The correlation is given by $\corr =
0.007\pm0.001$~GeV$^2$/$c^2$, which corresponds to the normalized dynamical 
fluctuation $\Sigma_{pt}$ of $\sim12\%$. 

\section{Conclusions and outlook}
In summary, the $D\overline{D}$ momentum correlations versus relative
azimuth in $p-p$ collisions at $\sqrt{s}$ = 14 TeV were investigated. 
The two-particle transverse momentum correlator 
${\langle\overline{\Delta}p_{t,1}\overline{\Delta}p_{t,2}\rangle}$ is a sensitive measure to carefully address these correlations.
 Our measure has a high sensitivity leading to strong $D\overline{D}$ 
back-to-back correlation and
helps to identify and disentangle different contributions to the
observed correlation pattern. 
 We demonstrated that the correlation of generated $c\overline{c}$ pairs survives the fragmentation process and even semileptonic decay to electrons (positrons) to 
a large extent. Thus, measurements of these correlations seem feasible with the upcoming collisions from LHC.




\begin{thebibliography}{9} 
\bibitem{strange} K.H. Ackermann {\it et al.} (STAR collaboration), Phys. Rev. Lett. {\bf 86}, 402 (2001); 
C. Adler {\it et al.} (STAR collaboration), Phys. Rev. Lett. {\bf 89}, 132301 (2002); 
J. Adams {\it et al.} (STAR collaboration), Phys. Rev. Lett. {\bf 95}, 122301 (2005); 
B.I. Abelev {\it et al.} (STAR collaboration), Phys. Rev. C {\bf 69}, 054907 (2004). 

\bibitem{thermalization} E.L. Bratkovskaya {\it et al.}, Phys. Rev. C {\bf 69}, 054907 (2004). 
\bibitem{Zhu} X. Zhu {\it et al.}, Phys. Lett. B {\bf 647}, 366 (2007).
\bibitem{pythia}T. Sj\"ostrand {\it et al.}, Comput. Phys. Commun {\bf 135}, 238 (2001). 
\bibitem{review}C. Louren\c{c}o and H.K. W\"ohri, Phys. Rept. {\bf 433}, 127 (2006). 
\bibitem{Dainese}N. Carrer and A. Dainese (ALICE Collaboration), arXiv:hep-ph/0311225.
\bibitem{pythia2}E. Norrbin and T. Sj\"ostrand, Eur. Phys. J. C {\bf 17}, 137 (2000). 
\bibitem{Zhu2} X. Zhu {\it et al.}, Phys. Rev. Lett. {\bf 100}, 152301 (2008).
\bibitem{pipj}J. Adams {\it et al.} (STAR collaboration), Phys. Rev. C {\bf 72}, 044902 (2005).
\bibitem{ceres-pt} D. Adamova {\it et al.} (CERES collaboration), Nucl. Phys. A {\bf 811}, 179 (2008).
\bibitem{spt}    D. Adamova {\it et al.} (CERES collaboration), Nucl. Phys. A {\bf 727}, 97 (2003).
\bibitem{spt1}H. Sako {\it et al.} (CERES collaboration), J. Phys. G {\bf 30}, S1371 (2004).
\bibitem{spt2}M. Rybczynski {\it et al.} (NA49 collaboration), J. Phys. Conf. Ser.  {\bf 5}, 74 (2005).
\bibitem{flow}E. Cuautle and G. Pai\'c, AIP Conf. Proc. {\bf 857}, 175 (2006).  

\end{thebibliography}
\begin{footnotesize}

\end{footnotesize}
\end{document}